# Time-Dependent Nuclear Energy-Density Functional Theory Toolkit for Neutron Star Crust: Dynamics of a Nucleus in a Neutron Superfluid

Daniel Pęcak,[1,2,*] Agata Zdanowicz,[1] Nicolas Chamel,[3,†] Piotr Magierski,[1,4,‡] and Gabriel Wlazłowski[1,4,§]

[1]Faculty of Physics, Warsaw University of Technology, Ulica Koszykowa 75, 00-662 Warsaw, Poland
[2]Institute of Physics, Polish Academy of Sciences, Aleja Lotnikow 32/46, PL-02668 Warsaw, Poland
[3]Institut d'Astronomie et d'Astrophysique, CP-226, Université Libre de Bruxelles, 1050 Brussels, Belgium
[4]Department of Physics, University of Washington, Seattle, Washington 98195-1560, USA

We present a new numerical tool designed to probe the dense layers of neutron star crusts. It is based on the time-dependent Hartree-Fock-Bogoliubov theory with generalized Skyrme nuclear energy-density functionals of the Brussels-Montreal family. We use it to study the time evolution of a nucleus accelerating through superfluid neutron medium in the inner crust of a neutron star. We extract an effective mass in the low velocity limit. We observe a threshold velocity and specify mechanisms of dissipation: phonon emission, Cooper pairs breaking, and vortex rings creation. These microscopic effects are of key importance for understanding various neutron star phenomena. Moreover, the mechanisms we describe are general and apply also to other fermionic superfluids interacting with obstacles like liquid helium or ultracold gases.

## I. INTRODUCTION

Neutron stars are the compact remnants formed in the furnace of supernova explosions from the gravitational collapse of the core of progenitor stars with a zero-age main sequence mass $8 \lesssim M \lesssim 30 M_\odot$ [1]. Initially very hot with temperatures reaching $\sim 10^{12}$ K, they rapidly cool down to $\sim 10^9$ K within days by releasing most of their energy in neutrinos (see, e.g., Refs. [2,3]). The interior of a neutron star is so dense that it is highly degenerate and is expected to undergo various quantum phase transitions, as observed in some terrestrial materials at low enough temperatures. In particular, neutrons present in the inner crust and in the core of a neutron star are thought to become superfluid by forming $^1S_0$ Cooper pairs, as in conventional superconductors. The existence of neutron superfluidity in neutron stars was predicted long ago before the actual discovery of these stars [4] and is now well established (see, e.g., Refs. [5,6]). In contrast to other superfluids that can be probed experimentally such as liquid helium or ultracold gases, nuclear superfluidity comes with much larger uncertainties. The conditions prevailing inside a neutron star are so extreme that they cannot be reproduced in the laboratory. In particular, the dynamical properties of nuclear superfluids raise many open questions concerning, for example, mutual entrainment between neutrons and protons, vortex dynamics, disspation, etc. While the properties of dense matter cannot be measured directly they can be probed indirectly through astrophysical observations. However, microscopic models of dense matter remain crucial for constructing global models of superfluid neutron stars [7]. The advances in computational physics open the possibility for supporting the development of global models through numerical experiments: advanced and accurate numerical modeling of various dynamical processes at the microscopic scale.

Among the most successful fully self-consistent approaches for modeling quantum systems is the density functional theory (DFT). It has become a standard theoretical tool in electronic systems, providing results at the same computational cost as mean-field calculations but with a higher precision; the error in the energy is typically below 1% [8,9]. In nuclear physics, the DFT (although we refer here to the DFT for both electronic and nuclear systems, some conceptual differences exist due to the breaking of symmetries in the latter [10–12]) has also been a method of choice for describing the properties of atomic nuclei across the whole nuclear chart and beyond, including nuclear matter under extreme astrophysical conditions (see the reviews Refs. [13–16]). At the same time, the DFT is very flexible. There are a variety of extensions of the formalism: for static and time-dependent problems [17–19], zero and finite-temperature problems [20–22], and for systems in normal and superconducting or superfluid states [23–28]. Implementing the DFT concept in the form of ready-to-use packages made it a workhorse for condensed-matter physics and quantum chemistry [29].

[*]Contact author: daniel.pecak@ifpan.edu.pl
[†]Contact author: nicolas.chamel@ulb.be
[‡]Contact author: piotrm@uw.edu
[§]Contact author: gabriel.wlazlowski@pw.edu.pl

In this work, we provide a tool for numerical explorations of dense-matter properties under conditions expected to be found inside the inner crust of neutron stars. The tool exploits opportunities offered by DFT. Our method relies on recent nuclear energy-density functional developments. Specifically, we use a family of Brussels-Montreal Skyrme (BSk) nuclear functionals that have been optimized for astrophysical applications [30]. These functionals provide a high-quality global description of various properties of finite nuclei (masses, radii, etc.) as well as properties of infinite nuclear matter in agreement with *ab initio* calculations. We combine them with techniques of high-performance computing (HPC) that recently reached an enormous scale, being able to perform the order of $10^{18}$ mathematical operations per second. As a result, we construct a general-purpose toolkit for 3D modeling of dynamical processes that take place inside neutron stars. The toolkit that hereafter we call W-BSk Toolkit [31] can deal with static and time-dependent phenomena at zero and finite temperatures without any symmetry constraints. The present HPC capabilities allow us to use it for modeling nuclear phenomena taking place in volumes exceeding $(100 \text{ fm})^3$—volumes that are sufficient to encapsulate hundreds of thousands of neutrons and protons. For example, one of the relevant scales when studying neutron star crust is the radius of the spherical Wigner-Seitz (WS) cell. Its precise value depends on the depth in the crust, and microscopic calculations yield values in the range $R_{\text{WS}} \in (10-60)$ fm [32]. The associated volumes of WS cell, $\frac{4}{3}\pi R_{\text{WS}}^3$, are within the reach of the toolkit; thus, nowadays, phenomena taking place on the scales of WS cells can be simulated by means of the microscopic self-consistent DFT approach. Superfluidity is crucial for describing nuclear systems when the dynamical properties are considered. Since the interactions between nucleons of the same species are naturally attractive at the densities of interest here, the formation of Cooper pairs is therefore unavoidable. The standard way of taking into account superfluid properties requires the usage of nonlocal order parameter [23,24]. Such an approach prevents its practical applications due to enormous numerical complexity. However, this was circumvented by formulating the problem using local pairing field [33] and the suitable framework had been developed (see review papers Refs. [15,34,35], and references therein). In this way, constructing an effective model of neutron star crust rooted in unified macroscopic formalism becomes feasible.

In the inner crust of a cold neutron star, matter is expected to exhibit various kinds of structures [36]. As a demonstration of the capabilities offered by the DFT framework, let us consider the range of average nucleon number densities for which protons and neutrons bind into quasispherical clusters immersed in a neutron superfluid bath; see Fig. 1. The description of the dynamical properties of such impurities poses a challenging theoretical problem, which requires us to describe properly the energy and momentum transfers between the impurities and the surrounding neutrons. There are some core similarities between the physics of the inner crust of neutron stars and other fields of physics, such as superfluid helium or ultracold gases. They can be used to anticipate phenomena occurring within the inner crust. While this is an excellent method for toy modeling, for example, the basic mechanism for glitches [37], it is not enough to make definite statements about properties of the inner crust due to major differences between nuclear and other systems. Therefore, the need for both a new paradigm and a tool to probe nuclear materials is highly important.

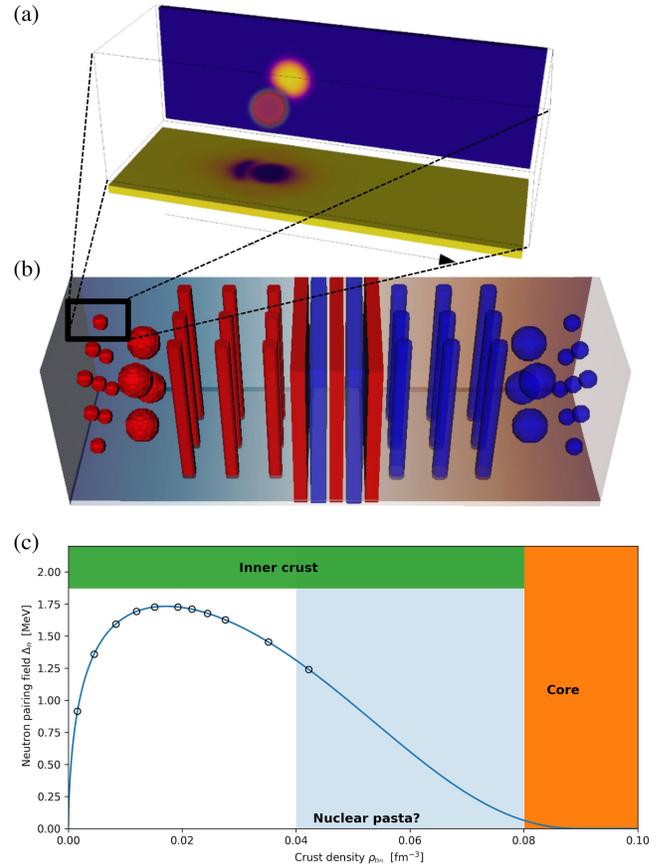

FIG. 1. In (a) we show the numerical setup considered in this work: a nucleus accelerating to the right along the $z$ axis through a neutron superfluid medium. The red sphere in the middle shows the proton density. The map at the bottom shows the neutron pairing field $\Delta_n(\mathbf{r})$ for $x = 0$ cut. The map beyond shows the neutron density $\rho_n(\mathbf{r})$ for $y = 0$ cut. (b) A schematic picture of the geometrical structures in the inner crust of a neutron star. Our region of interest corresponds to cases where protons form quasispherical self-bound impurities. At higher densities, exotic configurations such as rods or slabs referred to as "nuclear pasta" might be present. Red surfaces represent schematic boundaries of proton clusters in dilute neutron matter, while blue surfaces represent protonic holes in a dense nuclear matter. In (c) we show the $^1S_0$ neutron pairing gap $\Delta_n$ in neutron matter as a function of the background neutron density $\rho_{Bn}$ [38], on which the pairing part of the BSk functional is based. The circles indicate densities for which we extract the dynamical properties of the impurities.

One of the fundamental characteristics of an impurity is its inertia, whose modifications by the medium can be described by an effective mass. The problem of determining effective masses has in itself a long history, considered already by Landau and Pekar [39] and Fröhlich [40] in the context of electrons in solids. The situation considered here resembles the problem of a heavy impurity for which the general solution still evades theoretical description [41,42]. The goal is to integrate out the fermionic degrees of freedom and arrive at some effective equation of motion for the impurity alone. Such an equation should contain the change of the impurity mass by the medium and a dissipative term describing the irreversible energy flow between the impurity and the environment. This concept is crucial for polaron physics that has been recently explored experimentally within ultracold gasses platforms [43–45]. The nuclear impurity in the inner crust is made of protons tightly bound by nuclear forces and to which some neutrons are also attached. Most of those neutrons would be unbound in vacuum. They are held together with the proton cluster only because continuum states are already occupied. In other words, the impurity is closely connected to the surrounding neutron environment and would not exist without it. Moreover, the impurity is penetrable for neutrons. For these reasons, the effective characteristics of the impurity are not well established, and various predictions have been made. Most previous studies were carried out in the framework of classical hydrodynamics [46–50]. A quantum mechanical estimation of the effective mass of an impurity was first attempted in Ref. [51]. These calculations, however, were not self-consistent, and superfluidity was not explicitly taken into account. In all these cases, the motion of the impurity was assumed to remain stationary. This hypothesis only holds if the relative velocity of the impurity with respect to the neutron superfluid is sufficiently small. In reality, various kinds of perturbations can appear, including the nucleation of quantized vortices.

Here we demonstrate that the time-dependent DFT formalism provides a complementary method, which offers a new insight into the physical mechanisms associated with a moving impurity. These mechanisms are hard to describe within the static approach, especially for 3D geometry. For example, conceptually simple calculations of an impurity moving through the neutron superfluid deliver an abundance of physical information related to the effective mass of the impurity, the presence of dissipative forces, or even insight into the problem of the nucleation of quantum vortices in the interiors of neutron stars. Applications of the time-dependent DFT have been recently reported [52–54], but are limited to 1D structures (slabs). Here we present full 3D time-dependent DFT calculations without any symmetry restriction.

In Sec. II, we present the general theoretical framework, before describing, in Sec. III, the numerical experiment tailored to tackle the physical problem. The results are discussed in Sec. IV (effective mass) and Sec. V (dissipation channels). We conclude in Sec. VI.

## II. DFT AS GENERAL-PURPOSE FRAMEWORK FOR NUCLEAR MATTER

The popularity of DFT methods arises from a very good balance of the quality of predictions to the computation cost. To avoid prohibitive computing times, we consider only semilocal functionals, such as the Brussels-Montreal BSk functionals [30]. These functionals, which were specifically constructed for astrophysical applications, are based on generalized Skyrme effective interactions with density-dependent $t_1$ and $t_2$ terms [55] together with microscopically deduced contact pairing interaction [56–58]. The resulting partial differential equations, one needs to solve, have the same structure as the Hartree-Fock-Bogoliubov (HFB) equations with local fields. Their generic structure is (for brevity we omit position and time dependence)

$$i\hbar \frac{\partial}{\partial t} \begin{pmatrix} u_{k\uparrow} \\ u_{k\downarrow} \\ v_{k\uparrow} \\ v_{k\downarrow} \end{pmatrix} = \begin{pmatrix} h_{\uparrow\uparrow} & h_{\uparrow\downarrow} & 0 & \Delta \\ h_{\downarrow\uparrow} & h_{\downarrow\downarrow} & -\Delta & 0 \\ 0 & -\Delta^* & -h^*_{\uparrow\uparrow} & -h^*_{\uparrow\downarrow} \\ \Delta^* & 0 & -h^*_{\downarrow\uparrow} & -h^*_{\downarrow\downarrow} \end{pmatrix} \begin{pmatrix} u_{k\uparrow} \\ u_{k\downarrow} \\ v_{k\uparrow} \\ v_{k\downarrow} \end{pmatrix}, \quad (1)$$

where $[u_{k\uparrow}, u_{k\downarrow}, v_{k\uparrow}, v_{k\downarrow}]^T$ are four-component quasiparticle orbitals: mixtures of particles ($v_{k\sigma}$) and holes ($u_{k\sigma}$) with a set of quantum numbers $k$ and spin projections $\sigma = \{\uparrow, \downarrow\}$. The $h_{\sigma\sigma'}$ and $\Delta$ are single-particle Hamiltonian and pairing potential, respectively. In the case of nuclear problems, the terms $h_{\uparrow\downarrow}$ and $h_{\downarrow\uparrow}$ arise due to spin-orbit interaction. It is imperative in the description of nuclei; for example, it is responsible for the correct reproduction of nuclear magic numbers. On the other hand, the spin-orbit term does not contribute in homogeneous matter, and was shown to be very small in the crust of neutron stars [32]. On the technical level, the presence of $h_{\uparrow\downarrow}$ and $h_{\downarrow\uparrow}$ terms significantly increases the computation cost. There are numerical packages for nuclear dynamics, like LISE [59], that are optimized toward studies of nuclear reactions. We optimize our toolkit toward applications to neutron stars, and in this context, the ability to model phenomena in large volumes (compared to the size of nuclei in vacuum) is of major importance. Because of predicted small spatial variability of nuclear densities, the spin-orbit coupling is of secondary importance. We also assume that the nuclear matter is spin balanced and this condition is usually met, since the magnetic fields should be above $10^{17}$ G to polarize nuclear matter [60], while typically it is below $10^{15}$ G [61]. For these reasons, we neglect $h_{\downarrow\uparrow}$. The computation savings emerging from this simplification will allow us to consider volumes exceeding $100^3$ fm$^3$.

Under the assumptions $h_{\uparrow\downarrow} = h_{\downarrow\uparrow} = 0$ and $h_{\uparrow\uparrow} = h_{\downarrow\downarrow} = h$, it is sufficient to solve the HFB equations for a two-component vector [62],

$$i\hbar \frac{\partial}{\partial t} \begin{pmatrix} u_{q,k\uparrow} \\ v_{q,k\downarrow} \end{pmatrix} = \begin{pmatrix} h_q & \Delta_q \\ \Delta_q^* & -h_q^* \end{pmatrix} \begin{pmatrix} u_{q,k\uparrow} \\ v_{q,k\downarrow} \end{pmatrix}, \quad (2)$$

where we added an extra index $q = n, p$ to indicate that these equations must be solved for neutrons and protons, respectively. There is another set of equations for components $\{u_{q,k\downarrow}, v_{q,k\uparrow}\}$. However, there is no need to solve them independently, as the solution can be obtained from $\{u_{q,k\uparrow}, v_{q,k\downarrow}\}$ via suitable transformation [63].

To solve the time-dependent equations (2) one needs to provide an initial configuration specified by $\{u_{q,k\uparrow}(r,0), v_{q,k\downarrow}(r,0)\}$. It is usually obtained as a solution of static HFB equations (for clarity, we omit in notation the position dependence):

$$\begin{pmatrix} h_q & \Delta_q \\ \Delta_q^* & -h_q^* \end{pmatrix} \begin{pmatrix} u_{q,k\uparrow} \\ v_{q,k\downarrow} \end{pmatrix} = E_{q,k} \begin{pmatrix} u_{q,k\uparrow} \\ v_{q,k\downarrow} \end{pmatrix}. \quad (3)$$

Solving these equations self-consistently for the nonuniform matter in the inner crust of a neutron star, by itself, constitutes a challenging problem, and most of the results presented in the literature are limited to this type of calculations with the further approximation of a spherical WS cell, as, e.g., in Refs. [64–66].

The core of every DFT method is an energy-density functional $\mathcal{E}[\rho_q, \nu_q, \tau_q, j_q]$ that depends on the normal densities ($\rho_q$), anomalous densities ($\nu_q$), kinetic densities ($\tau_q$), currents ($j_q$), and in our case also on the gradients of density $\nabla \rho_q$. The densities are constructed from the quasiparticle states:

$$\rho_q = 2\sum_k [|v_{q,k\downarrow}|^2 f_T(-E_{q,k}) + |u_{q,k\uparrow}|^2 f_T(E_{q,k})], \quad (4)$$

$$\tau_q = 2\sum_k [|\nabla v_{q,k\downarrow}|^2 f_T(-E_{q,k}) + |\nabla u_{q,k\uparrow}|^2 f_T(E_{q,k})], \quad (5)$$

$$\nu_q = 2\sum_k u_{q,k\uparrow} v_{q,k\downarrow}^* [f_T(-E_{q,k}) - f_T(E_{q,k})], \quad (6)$$

$$j_q = 2\sum_k \text{Im}(v_{q,k\downarrow} \nabla v_{q,k\downarrow}^*) f_T(-E_{q,k})$$
$$+ 2\sum_k \text{Im}(u_{q,k\uparrow} \nabla u_{q,k\uparrow}^*) f_T(E_{q,k}). \quad (7)$$

The factor 2 in the above equations comes from the spin degeneracy of neutrons and protons. The summation over $k$ is restricted only to the interval below the cutoff energy, $0 < E_{q,k} < E_{\text{cut}}$. The thermal occupation factors ($k_B$ is Boltzmann constant),

$$f_T(E) = \left[1 + \exp\left(\frac{E}{k_B T}\right)\right]^{-1}, \quad (8)$$

have been introduced to `W-BSk Toolkit` to allow investigations at finite temperatures $T$. With these factors the framework becomes formally equivalent to the finite-temperature HFB method [67].

The energy of the system has a generic form:

$$E = \int \mathcal{E}[\rho_q, \nu_q, \tau_q, j_q] dr - \sum_{q=n,p} \int \left(\mu_q - V_q^{(\text{ext})}\right) \rho_q dr$$
$$- \frac{1}{2} \sum_{q=n,p} \int \left(\Delta_q^{(\text{ext})} \nu_q^* + \Delta_q^{(\text{ext})*} \nu_q\right) dr$$
$$- \hbar \sum_{q=n,p} \int v_q^{(\text{ext})} \cdot j_q dr. \quad (9)$$

The first term in the above expression is the intrinsic energy of the system. The remaining three terms are introduced as generalized Lagrange multipliers. In general, external field $V_q^{(\text{ext})}$, external pairing field $\Delta_q^{(\text{ext})}$, and external velocity field $v_q^{(\text{ext})}$ are position and time dependent. They are coupled to normal $\rho_q$, anomalous $\nu_q$, and current $j_q$ densities, respectively, and allow us to control them depending on the problem considered. Note that from the $V_q^{(\text{ext})}$, we have extracted the constant part $\mu_q$ that has the meaning of the chemical potential. If $V_q^{(\text{ext})}(r,t) = 0$, one recognizes that the associated term reduces to the well-known form $-\mu_q N_q$, where the number of particles is given by $N_q = \int \rho_q(r,t) dr$. The introduction of generalized Lagrange multipliers makes the toolkit applicable to a plethora of problems. For example, setting $v_q^{(\text{ext})} = \Omega \times r$ results in the requirement of putting our system into rotation with angular velocity $\Omega$. In this work, we will use the external potential $V_q^{(\text{ext})}(r,t)$ to simulate the presence of an external electric field.

In this particular study, we adopted the energy-density functional BSk31 [30]. This functional from the Brussels-Montreal family was constructed from generalized Skyrme effective interactions [55] and has the following form:

$$\mathcal{E}[\rho_q, \nabla \rho_q, \nu_q, \tau_q, j_q]$$
$$= \frac{\hbar^2}{2m_n} \tau_n + \frac{\hbar^2}{2m_p} \tau_p + \mathcal{E}_\tau(\rho_q, \tau_q, j_q) \quad (10\text{a})$$
$$+ \mathcal{E}_\rho(\rho_q) + \mathcal{E}_{\Delta\rho}(\rho_q, \nabla \rho_q) \quad (10\text{b})$$
$$+ \mathcal{E}_\pi(\rho_q, \nabla \rho_q, \nu_q) + \mathcal{E}_{\text{Coul}}(\rho_p). \quad (10\text{c})$$

The different terms above have the following meaning. In Eq. (10a), the first two terms correspond to the kinetic energy density of protons and neutrons, while $\mathcal{E}_\tau$ is the energy density arising from the momentum-dependent part of the effective interaction. This last term is responsible for the density-dependent effective mass of the nucleons (in contrast to the effective mass of the impurity that we study in this paper) and for the mutual neutron-proton entrainment effects due to current-current couplings [68,69]. The two terms in Eq. (10b) contribute to the mean-field potential felt by a nucleon due to its interactions with the dense background and its spatial fluctuations. The last two terms in Eq. (10c) account for pairing (see Appendix A) and Coulomb interactions.

The single-particle mean fields $h_q$ and pairing potentials $\Delta_q$ are defined for a given density functional $\mathcal{E}$ through variation over proper densities:

$$h_q = -\boldsymbol{\nabla}\frac{\delta\mathcal{E}}{\delta\tau_q}\boldsymbol{\nabla} + \frac{\delta\mathcal{E}}{\delta\rho_q} - (\mu_q - V_q^{(\text{ext})}) - \frac{i}{2}\left\{\frac{\delta\mathcal{E}}{\delta\boldsymbol{j}_q} - \hbar\boldsymbol{v}_q^{(\text{ext})}, \boldsymbol{\nabla}\right\}, \quad (11)$$

$$\Delta_q = -2\frac{\delta\mathcal{E}}{\delta\nu_q^*} + \Delta_q^{(\text{ext})}. \quad (12)$$

According to our shorthand notation the term $\delta\mathcal{E}/\delta\boldsymbol{j}_q$ describes components $x$, $y$, $z$ arising from the variation of the energy with respect to the current density components $j_{qx}$, $j_{qy}$, $j_{qz}$, respectively. We denote anticommutator by $\{\cdot,\cdot\}$. One can now recognize the mathematical complexity of the problem: The potentials $h_q$ and $\Delta_q$ depend on densities Eqs. (4)–(7), which in turn are expressed through the quasiparticle states $\{u_{q,k\uparrow}, v_{q,k\downarrow}\}$. They need to be obtained from Eqs. (2) and (3) that are defined via $h_q$ and $\Delta_q$.

To tackle the problem, we designed a software W-BSk Toolkit. It is based on the numerical engine of W-SLDA Toolkit [70], constructed for ultracold atomic gases. The W-SLDA engine has already been applied to various problems. The landscape of problems encapsulates quantum vortices [71–74], solitons [75], exotic states in spin-imbalanced gases [76–79], Josephson junctions [80], Higgs modes [81], and quantum turbulence [82,83]. In the first stage, we adapted the code for pure neutron matter and applied it to investigations of quantum vortices at finite temperatures [84] using the functional BSk31 [30]. Here, we accomplished the second stage, extending it to take into account the presence of protons. In the present study, we use the same functional BSk31. We also assume the whole system is electrically charge neutral; the positive proton charge is compensated by the presence of a uniform electron gas. The full expression of the functional $\mathcal{E}$ and the associated mean fields can be found in Appendix A.

Cartesian lattice is used to discretize the space in Eqs. (2) and (3). Consequently, the quasiparticle wave functions are expressed on a mesh of size $N_x \times N_y \times N_z$ with lattice spacings $\Delta x$, $\Delta y$, and $\Delta z$, respectively. Tests indicated that $\Delta x = \Delta y = \Delta z \approx 1.25$ fm provides already high-quality representation [59,85]. The equations are solved imposing periodic boundary conditions. The lattice spacing provides a natural energy cutoff scale $E_{\text{cut}} = \hbar^2\pi^2/(2m_n\Delta x^2)$. It is located in energy window $E_{\text{cut}} \approx (130–200)$ MeV for lattice spacings spanning (1.00–1.25) fm, which is significantly larger than the Fermi energy of nucleons. That implies that the number of considered states is also much larger than the number of simulated particles. The large values of the energy cutoff are required in order to maintain the trustable evolution of the nuclear system [15,86,87]. The lattice representation allows for efficient utilization of spectral methods for computation of derivatives [88]. These are high-accuracy methods with the associated numerical cost comparable to finite-difference methods (precisely, the computation cost is set by the efficiency of a fast Fourier transform). The time integration of Eq. (2) is done via the multistep Adams-Bashforth-Moulton method of the fifth order. This integrator generates stable and accurate length trajectories $\sim 10^4$ fm/$c$. The robustness of the time integrator with respect to the perturbations of the initial states has been demonstrated in Ref. [89]. The static problem Eq. (3) is solved by a series of direct diagonalizations supported by the ELPA library [90–92]. Convergence of the self-consistent loop is greatly improved by utilization of the Broyden algorithm [93]. The solvers for static and time-dependent problems are parallelized by message passing interface (MPI) protocol, and graphics processing units (GPUs) are used to achieve the highest computation performance. The implementation supports both nVidia and AMD accelerators, which are the two major brands of GPUs. For example, when running on tier-0 systems, like LUMI [94], it can deal with problems formulated on the lattice of size $100^3$ [95]. However, it can also be used on smaller computing systems, depending on the problem size and type. For example, the static calculations presented here (as shown in Fig. 2) were performed with relatively small resources. The implementation is released in the form of an open-source code. It is accessible via the Web page Ref. [31], so everyone can inspect the implementation and reuse solutions derived within this W-BSk Toolkit project.

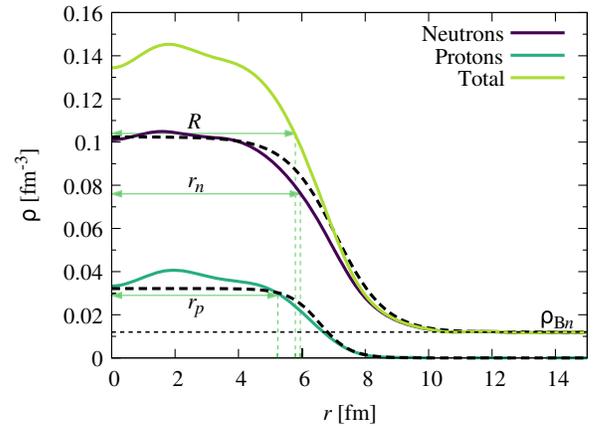

FIG. 2. Typical density profile for a $_{40}$Zr cluster existing in the layer of the inner crust of a neutron star at the average nucleon density $\bar{\rho} = 0.01481$ fm$^{-3}$. Far from the cluster, neutrons are uniformly distributed with a density $\rho_{\text{B}n} = 0.01196$ fm$^{-3}$. Solutions of the HFB equations (solid lines) are compared with fourth-order extended Thomas-Fermi calculations (dashed lines) [96]. By $r_p$ and $r_n$ we denote the root-mean-square radii of protons and neutrons (after subtracting the neutron background density).

## III. NUMERICAL EXPERIMENT

We consider a small matter element of the inner crust: a nuclear cluster immersed in a sea of superfluid neutrons at zero temperature $T \to 0$; see Fig. 1(a). The most stable clusters expected to be present in broad regions of the inner crust are very neutron-rich zirconium isotopes with proton number $Z = 40$ [96]. In order to investigate its dynamical properties, we accelerate it through the superfluid medium by applying a constant electric field in a $z$ direction, $\boldsymbol{E} = [0, 0, E_z]$. As a result, protons move due to the electric force $\boldsymbol{F} = Ze\boldsymbol{E}$, dragging a certain number of neutrons. Technically, the constant force is modeled by a linear external potential that couples only to protons:

$$V_p^{(\text{ext})}(\boldsymbol{r}) = -\frac{1}{Z}\boldsymbol{F} \cdot \boldsymbol{r}. \quad (13)$$

The numerical simulation consists of three main steps: (i) finding a self-consistent solution for the ground state of the cluster, (ii) extracting the initial wave function $\{u_{q,k\uparrow}(\boldsymbol{r}, 0), v_{q,k\downarrow}(\boldsymbol{r}, 0)\}$, and (iii) evolving the system in time. In the first step, we solve self-consistently the HFB equations (3) with the constraints $N_p = Z = 40$ and the density of neutrons far from the nucleus $\rho_{\text{B}n}$. This is achieved by properly adjusting the chemical potentials $\mu_q$. We solve the equations without any geometrical restriction on a uniform cubic grid $32 \times 32 \times 32$ with a lattice spacing $\Delta x = \Delta y = \Delta z = 1.25$ fm. The resulting volume $40^3$ fm$^3$ is large enough to saturate the values of bulk neutron density to the desired value $\rho_{\text{B}n}$ that corresponds to the layer in the inner crust with average density $\bar{\rho}$ (see Table I). The selected values of $\rho_{\text{B}n}$ are indicated by circles in Fig. 1(c). The algorithm converges to a spherically symmetric distribution of protons and neutrons in all considered cases. Examples of 1D density distributions obtained for $\rho_{\text{B}n} = 0.01196$ fm$^{-3}$ are shown in Fig. 2. We compare our results with semiclassical predictions represented by dashed black lines and based on fourth-order extended Thomas-Fermi calculations in spherical Wigner-Seitz cells [96]. In this approach, the shell and pairing effects were neglected and the nucleon-density distributions were parametrized; thus, the density distributions are almost flat inside the impurity, contrary to self-consistent calculations.

From the static calculations, we may estimate the number of bound neutrons as follows:

$$N_{\text{bound}} = \int [\rho_n(\boldsymbol{r}) - \rho_{\text{B}n}] d\boldsymbol{r}. \quad (14)$$

This definition gives the proper result in the limit $\rho_{\text{B}n} \to 0$ (the nucleus in the vacuum), and it provides an estimate for the effective mass of the impurity according to the formula

$$M_{\text{eff}}^{(s)} = Zm_p + N_{\text{bound}}m_n, \quad (15)$$

where $m_p$ and $m_n$ are masses of protons and neutrons, respectively. In the calculations, we assume that $m_p \approx m_n \approx 939.57$ MeV/$c^2$. To determine the size of the impurity we use the root-mean-square radius,

$$R = \sqrt{\langle R^2 \rangle} = \sqrt{\frac{\int \rho_{\text{tot}}(\boldsymbol{r}) r^2 d\boldsymbol{r}}{\int \rho_{\text{tot}}(\boldsymbol{r}) d\boldsymbol{r}}}, \quad (16)$$

where $\rho_{\text{tot}}(\boldsymbol{r}) = \rho_n(\boldsymbol{r}) - \rho_{\text{B}n} + \rho_p(\boldsymbol{r})$. Similarly, we can define the neutron and proton root-mean-square radii associated with bound neutrons $r_n$ and protons $r_p$, using the densities $\rho_n(\boldsymbol{r}) - \rho_{\text{B}n}$ and $\rho_p(\boldsymbol{r})$, respectively. These quantities are depicted in Fig. 2. The tests provide satisfactory agreement with the results reported in Ref. [96]. Note that the calculations presented here, apart from neglecting the spin-orbit coupling and the gradient pairing term (see Appendix A), contain all contributions to the energy-density functional (nuclear, pairing, and Coulomb terms) taken into account in a fully self-consistent manner (there are no other hidden approximations). The good agreement with the independent calculations demonstrates the correctness of the energy functional implementation within our toolkit.

Next, in stage (ii), we triple the volume of the system by adding two cubes (each of volume $40^3$ fm$^3$) filled with uniform superfluid neutron matter of density $\rho_{\text{B}n}$ to the side of the solution from stage (i). In this way, we prepare a sufficient space for the motion of the impurity, which will be performed in volume $40^2 \times 120$ fm$^3$. (The tools allowing for the manipulations of static solutions are also included in the toolkit.) Note that one does not need a very accurately converged ground state for the time evolution procedure. Since we are interested in nonequilibrium dynamics, it is enough to be fairly close to the ground state, i.e., within an excitation energy that is small compared to excitation energies generated by the dynamics. Therefore, in stage (ii) it is sufficient to perform just a few iterations of the self-consistent method, which at this stage becomes costly since we are now diagonalizing HFB matrices of size $2 \times 32^2 \times 96 = 196\,608$. (We estimated that the energy is minimized up to the accuracy of the order of keV per nucleon.) At the end of this process, the wave functions $\{u_{q,k\uparrow}(r,0), v_{q,k\downarrow}(r,0)\}$ are stored. The total number of states (for protons and neutrons) in energy interval $E_{q,k} \in [0, E_{cut} = 130 \text{ MeV}]$ exceeds slightly a hundred thousand, while total number of nucleons is from 456 for the lowest ($\rho_{Bn} \approx 0.002$ fm$^{-3}$) up to 8150 for the highest ($\rho_{Bn} \approx 0.051$ fm$^{-3}$) densities.

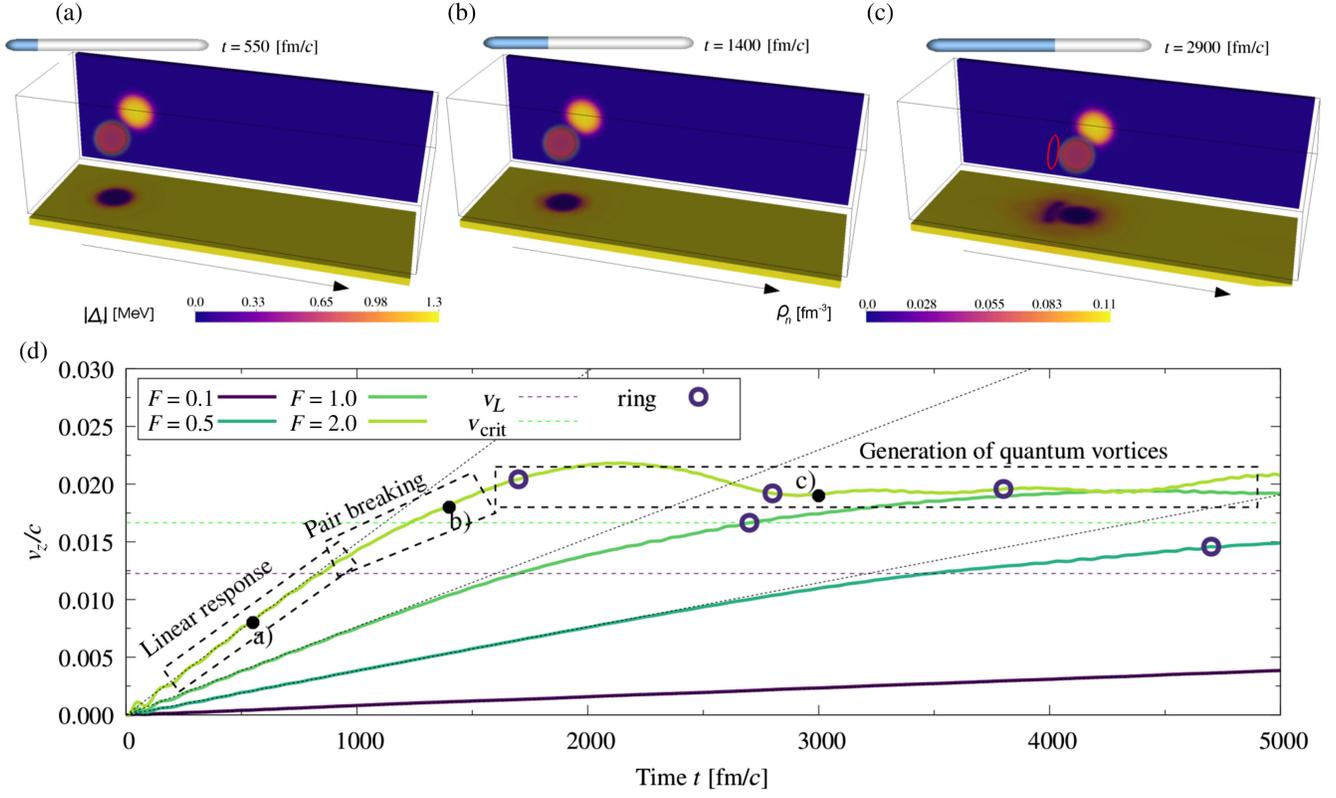

FIG. 3. (a)–(c) Snapshots of the impurity dynamics for selected times (550, 1400, and 2900 fm/c) as indicated in (d). Each frame shows proton density (red sphere in the box), neutron density cross section $\rho_n$ (behind), modulus of the neutron pairing field $|\Delta_n|$ (bottom). The black arrow indicates the direction of a constant electric field. See the dataset [97] for the full movie. (d) The velocity of the center of mass of protons dragged in superfluid neutrons of density $\rho_{Bn} \approx 0.0045$ fm$^{-3}$ for different forces $F$ (measured in MeV/fm) as a function of time in fm/c. For this density, we observe three distinct regimes of dynamics: linear response, dissipative dynamics induced by the breaking of Cooper pairs, and quantum vortex proliferation for certain moments (denoted by blue circles). The vortices are created for velocities above the Landau velocity $v_L$ (blue dashed line). Impurity velocities tend to the critical velocity $v_{crit}$ (green dashed line). Black dotted lines show the linear fit for the short duration of dynamics.

The resulting quantum states are subsequently evolved [step (iii)], using the time-dependent HFB equation (2). At this stage, we introduce a constant electric field along the longest side of the simulation box, that we denote as $z$, via Eq. (13). We turn on the electric field $E_z(t)$ gradually within the time interval $\Delta t = 10$ fm/c (for the explicit protocol, see Appendix B). The equations of motion are integrated with a time step $dt = 0.1$ fm/c. We monitor the conservation of the particle number and the total energy (after the external electric field is turned on; see Appendix B), which is satisfied with high accuracy (the particle number is conserved within a precision of $10^{-6}\%$ and the total energy with a precision from 0.5% up to 0.001% over the trajectory length $\Delta t = 5000$ fm/c, depending on the density). The simulations were executed for different strengths of $F \equiv ZeE_z$. Sample frames from simulations are shown in Figs. 1(a) and 3(a)–3(c). All other technical settings are provided in the dataset [97].

The real-time dynamics provides insight into the evolution of the densities Eqs. (4)–(7). From them, we can obtain the desired observables. The time evolution of the position of the center of mass of protons,

$$\boldsymbol{R}_{c.m.}(t) = \frac{1}{Z} \int \rho_p(r,t) r \, dr, \qquad (17)$$

immediately reveals the existence of different dynamical regimes. In Fig. 3(d), we show the $z$ component of the velocity of the center of mass (the velocity along the $x$ and $y$ directions is zero),

$$\boldsymbol{v}_{c.m.}(t) = \frac{d\boldsymbol{R}_{c.m.}(t)}{dt} = [0, 0, v_z(t)], \qquad (18)$$

for a few values of external forces $F$. As long as the velocity is not too large, it changes linearly with time $v_z(t) \sim a_z t$, except for the initial stage of the evolution, discussed in Appendix C, which is due to the excitation of the collective mode being an analog of isovector giant dipole resonance (IGDR). Apart from the effect of IGDR excitation, which can be minimized by decreasing the rate of switching on the electric field, the regime is essentially dissipationless, and the impurity moves according to Newton's law, $F = M_{\text{eff}} a_z$. Since $F$ is the control parameter and $a_z$ can be accurately extracted from the simulated data, we obtain an unequivocal method of extracting the effective mass. Once the impurity is accelerated above a threshold value (which is density dependent), the constant external force no longer induces a uniformly accelerated motion. This signals that additional forces start to act on the impurity. This regime provides insight into dissipative effects, due to irreversible coupling of the impurity with the superfluid bath. These conceptually simple numerical experiments involving time evolution turn out to provide a plethora of information that is either hard or impossible to extract from static calculations.

## IV. EFFECTIVE MASS

For all considered densities, we found that, initially, the impurity moves with a constant acceleration, irrespective of the applied constant force $F$. The initial linear increase of the impurity velocity is a consequence of the superfluidity. In the regime where the superfluid velocity does not exceed Landau's critical velocity, no quasiparticle can be excited and the motion is dissipationless [98,99]. This peculiar feature is crucial for the successful extraction of the effective mass. Namely, the presence of an energy gap in the excitation spectrum of the system makes the determination of effective mass a well-posed question in contrast to the normal fluids, where particle-hole excitations occur at arbitrary small velocity, making the problem of disentangling reversible and irreversible energy exchange between impurity and environment practically impossible. In Fig. 3(d), we plot the velocity of the center of mass of protons extracted from the simulations, together with fitted linear functions $a_z t$ for the initial part of the trajectory. As expected, the ratio $M_{\text{eff}}^{(d)} = a_z/F$ turns out to be almost independent of the force. We show the fitted values of effective mass $M_{\text{eff}}^{(d)}$ for various densities in Fig. 4 as blue diamonds. The error bars indicate the differences of effective mass $M_{\text{eff}}^{(d)}$ extracted for different magnitudes of the force accelerating the nucleus. The effective mass is significantly higher than $Zm_p$, and the difference $M_{\text{eff}}^{(d)} - Zm_p$ is related to the number of neutrons that are effectively bound (including those entrained by protons).

The extracted effective mass $M_{\text{eff}}^{(d)}$ can be compared to other approaches. The simplest one, and most commonly adopted, relies on the formula (15). The key idea is to distinguish between bound and free neutrons based on the spatial properties of the single-particle density distribution. This approach has semiclassical roots as it provides a reasonable answer in the limit of $k_{\text{F}n} R \to \infty$, where the Fermi wave vector is given by $k_{\text{F}n} = (3\pi^2 \rho_n)^{1/3}$. We also assume that the length scale of spatial variations of density is large compared to the Fermi wavelength. In such a case, contributions coming from shell effects and resonant states can be neglected. Both these assumptions can be questioned in the context of neutron star crust. The other approach is based on hydrodynamic considerations. The modification of the mass of a moving impurity comes from the excitation of the flow in the superfluid medium, which is assumed to be incompressible and irrotational. Under such assumptions, the flow that results merely from mass conservation can be easily extracted. Namely, if the impurity is chosen to be a sphere of radius $R_h$ with a sharp surface filled with fluid of density $\rho_{\text{in}}$, whereas the outside density of fluid $\rho_{\text{out}}$ is associated with the bulk neutrons, then the energy cost of exciting the superfluid flow scales with the velocity $V$ as $E \sim V^2$. The proportionality coefficient quantifies the effective mass, and according to Ref. [48] it reads

$$M_{\text{eff}}^{(h)} = \frac{4}{3}\pi R_h^3 m_n \frac{(\rho_{\text{in}} - \rho_{\text{out}})^2}{\rho_{\text{in}} + 2\rho_{\text{out}}}. \quad (19)$$

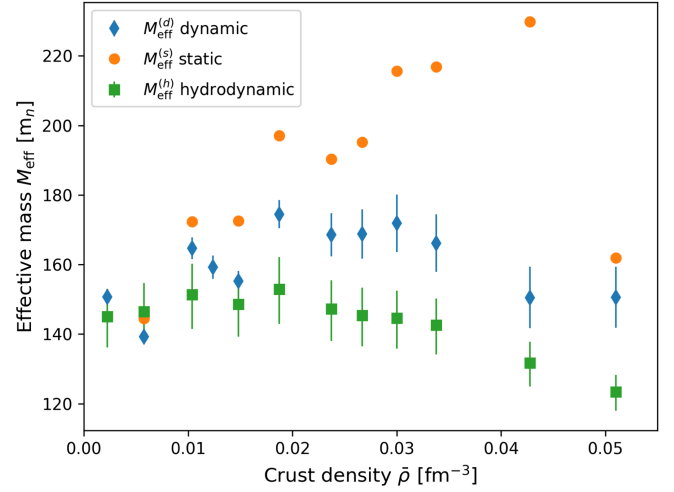

FIG. 4. The effective mass of the impurity calculated with different approaches: dynamic $M_{\text{eff}}^{(d)}$, static $M_{\text{eff}}^{(s)}$, hydrodynamic $M_{\text{eff}}^{(h)}$. The details, including how we determine the error bars, are explained in the main text. Static and dynamical calculations coincide in the low-density regime. We shifted the results given by hydrodynamic approximation to match the low-density limit.

One can see that the definition Eq. (19) simplifies in two important limits. In the absence of impurity (the matter is uniform $\rho_{\text{in}} = \rho_{\text{out}}$), the effective mass is zero. On the other hand, when the impurity is in vacuum ($\rho_{\text{out}} = 0$ and $\rho_{\text{in}}$ equals nuclear matter saturation density $\rho_0 \approx 0.16 \text{ fm}^{-3}$) the effective mass reduces to the "bare" mass obtained from the product of the nucleon mass by number of nucleons. To compare this formula with the effective mass extracted from time-dependent DFT we have identified densities $\rho_{\rm in}$, $\rho_{\rm out}$ as densities at the center of the impurity and the corner of the numerical box, respectively. Estimating $R_h$ using the root-mean-square radius $R$ as given by Eq. (16) underestimates the effective mass, resulting in $M_{\rm eff}^{(h)} < M_{\rm eff}^{(s)}$. In the vacuum limit, both approaches should give the same value of the effective mass. Therefore, we compensate the systematic error of determining $R_h$ by shifting all results for hydrodynamic approach by a constant value to meet the low-density limit. Since the surface of the realistic cluster is not sharp, we have estimated the uncertainty of the effective mass estimation within the hydrodynamics approach from the values corresponding to $95\%R_h$ and $105\%R_h$.

In Fig. 4 we compare the effective masses obtained within different approaches, together with their uncertainties. The dynamical approach based on time-dependent DFT does not rely on any assumption concerning the size and shape of the impurity. As expected, the static prescription $M_{\rm eff}^{(s)}$ matches the dynamical ones in the regimes of low densities $\bar{\rho} \lesssim 0.01$ fm$^{-3}$. At higher densities, $\bar{\rho} \gtrsim 0.03$ fm$^{-3}$, discrepancy can be observed. In static calculations, the size of the impurity and therefore also its effective mass grows with density, while in the dynamical approach the effective mass tends to saturate. This indicates that the simple division between bound and unbound neutrons, based on geometrical arguments only, becomes questionable. In contrast, the hydrodynamic approach qualitatively reproduces the behavior of the dynamical calculations, even though it systematically underestimates the effective mass for densities above $\bar{\rho} \gtrsim 0.02$ fm$^{-3}$. This can be understood from the fact that the hydrodynamics approach takes into account only one effect related to the superfluid motion, namely, the excitation of incompressible and irrotational flow. Note also that the applicability of the hydrodynamics approach can be questioned, if we recall that the impurity size $R$ is of the same order as the most important length scales of the neutron superfluid, namely, the coherence length $\xi = \hbar k_{Fn}/(\pi \Delta_n m_n)$. One would need $\xi \ll R$ for hydrodynamics to be quantitatively accurate.

The effective mass of a nuclear cluster moving in a neutron superfluid is a central issue for determining various properties of the inner crust of a neutron star. It alters the Coulomb crystal's dynamical properties. For example, the associated plasma frequency of the system is inversely proportional to the square root of the effective mass and reads $\omega_p = \sqrt{4\pi n_{\rm ion} Z^2 e^2/M_{\rm eff}}$, where $n_{\rm ion}$ is the ion density. Consequently, the crystal's phonon spectrum is modified, and the Coulomb crystal's thermodynamic functions will be modified as well [100]. Determination of the effective mass is also important for the thermal and electric conductivities of the inner crust, governed by electron-phonon scattering [101]. The effective mass of an impurity moving in a dense nucleon background has been also recently shown to play an important role in the formation of clusters in a hot newly born neutron star [102]. The clusterization of matter depends not only on surface and Coulomb effects but also on the (uncorrelated) translational motion of clusters, which in turn is modified by the medium through the renormalization of their mass.

## V. DISSIPATION

One of the hallmarks of a superfluid state is the existence of dissipationless flow. A typical example is an obstacle moving in a superfluid without creating any excitation. The above statement is correct under the assumption of zero temperature $T = 0$ and sufficiently small velocities $v < v_L$ of moving obstacles. This characteristic threshold velocity $v_L$ is the so-called Landau velocity. For a uniform neutron superfluid of density $\rho_n = k_{Fn}^3/(3\pi^2)$, it is approximately given by the well-known formula from the BCS theory [103] (see Ref. [98] for the exact treatment in the nuclear context),

$$v_L = \frac{\Delta_n}{\hbar k_{Fn}}. \quad (20)$$

At this velocity, there is enough kinetic energy to break a Cooper pair or equivalently excite a quasiparticle above the energy gap. In Fig. 3, we see that the uniformly accelerated motion is lost once the impurity velocity approaches $v_L$. The comparison for various densities is shown in Fig. 5(a).

To demonstrate that the Cooper pair breaking is responsible for the dissipative dynamics once $v_L$ is reached, let us consider the condensation energy of neutron Cooper pairs defined as

$$E_{\rm cond}(t) = \int \frac{3}{8} \frac{|\Delta_n(\boldsymbol{r},t)|^2}{\varepsilon_{Fn}(\boldsymbol{r},t)} \rho_n(\boldsymbol{r},t) d\boldsymbol{r}, \quad (21)$$

where $\varepsilon_{Fn}(\boldsymbol{r},t) = \hbar^2 [3\pi^2 \rho_n(\boldsymbol{r},t)]^{2/3}/[2m_n^*(\boldsymbol{r},t)]$ is the local Fermi energy of neutrons, and $m_n^*$ is effective mass of a neutron. In the case of a uniform system, it reduces to $3\Delta_n^2 N_n/(8\varepsilon_{Fn})$, which is a standard formula known from the BCS theory. The quantity Eq. (21) is displayed in Fig. 5(b). For velocities $v_z < v_L$, the velocity of the impurity increases linearly while the condensation energy remains fairly constant. Once the Landau velocity is reached, the condensation energy starts to drop. It is worth noting that the onset of Cooper pair breaking does not match exactly the velocity $v_L$. In general, one needs to consider the velocity field of neutrons. It can be computed by noting that $\boldsymbol{j}_n(\boldsymbol{r},t) = \rho_n(\boldsymbol{r},t)\boldsymbol{v}_n(\boldsymbol{r},t)$. The extracted $\boldsymbol{v}_n(\boldsymbol{r},t)$ at three selected time instants is displayed in Figs. 6(a)–6(c). The blue color shows regions with a velocity field below the threshold value $v_n(r) < v_L$, white color indicate regions where it becomes comparable to $v_n(r) \approx v_L$, and the red color shows regions where the velocity exceeds $v_\mathrm{L}$. The figure shows clearly that the process of destroying the Cooper pairs takes place in the vicinity of the impurity.

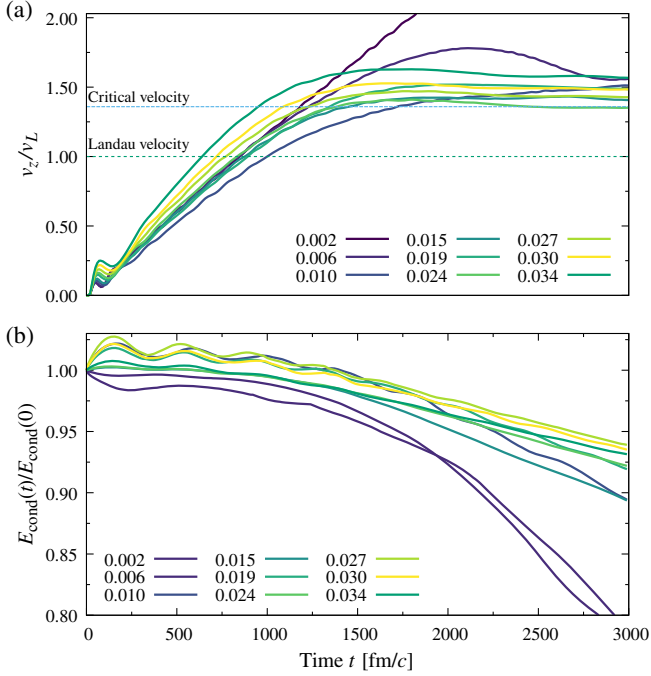

FIG. 5. (a) The velocity of the impurity as a function of time for different densities in fm$^{-3}$, normalized to the respective values of the Landau velocity. The data series corresponds to force $F = 2$ MeV/fm. By dashed and dotted lines, we indicate the two characteristic velocity scales given by Eqs. (20) and (22), respectively. (b) The condensation energy $E_\mathrm{cond}$ relative to its initial value as a function of time for the same data series as in (a).

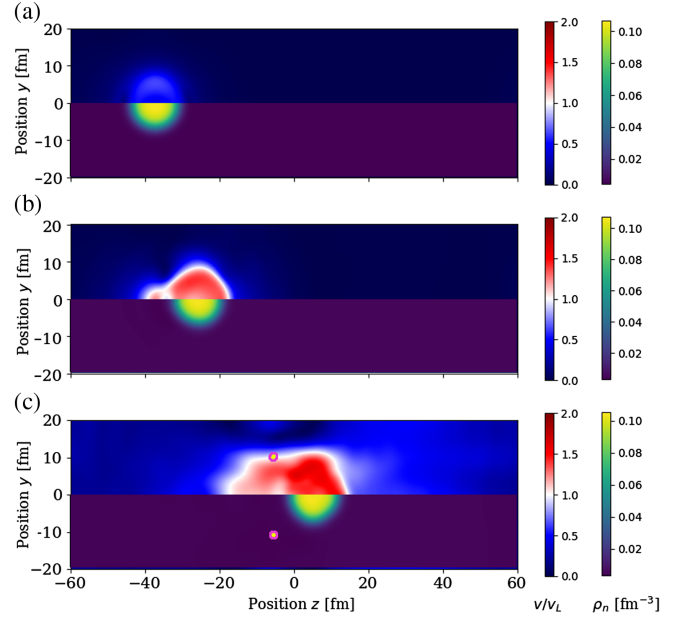

FIG. 6. Each panel presents the neutron density cross section through $x = 0$ (lower part) and local velocity in units of bulk Landau velocity (upper part). The consecutive panels are taken at times 550, 1400, and 2900 fm/c, which correspond to Figs. 3(a)–3(c). (a) In the linear response regime mainly the impurity is moving. (b) In the breaking pair regime the free neutrons in the vicinity of impurity are affected. (c) In the turbulent regime a large volume of neutrons is affected. Two points shown behind the impurity (at $z \approx -5$ fm) are the cross section of the vortex ring generated in this regime.

The stationary flows of nuclear homogeneous superfluids have been recently studied in Ref. [98] within the DFT approach. The superfluid has been shown to become gapless when the superfluid velocity exceeds Landau's velocity $v_\mathrm{L}$. It is a regime in which a normal fluid of quasiparticle excitations coexists with the superfluid, even at zero temperature [99]. Superfluidity is destroyed at a higher critical velocity given by

$$v_\mathrm{crit} = \frac{\exp(1)}{2} v_\mathrm{L} \approx 1.4 v_\mathrm{L}. \qquad (22)$$

Although the system considered here is not homogeneous and the flow is not stationary, our numerical simulations indicate that $v_\mathrm{crit}$ remains the characteristic velocity scale. In almost all cases, the speed of the impurity is indeed limited by $v_\mathrm{crit}$, as can be seen in Fig. 5(a). However, the normal component appears beyond $v_\mathrm{L}$ in the form of a vortex ring around the moving nucleus, as can be seen in Fig. 3(c). The exceptional behavior of the lowest density $\rho_{\mathrm{B}n} = 0.002$ comes from the fact that it is close to the limit of an impurity moving in vacuum, where the coupling to the environment is not expected.

The vortex-shedding process has been observed experimentally in superfluid He [104,105] and ultracold atomic gases [106,107]. It was also simulated numerically using the Gross-Pitaevskii equation (see examples reported in Refs. [108–110]). In these studies, the obstacle was impenetrable and much bigger than the coherence length. In the case of nuclear impurity, the situation is different. First of all, there is no clear separation of scales. Second, the impurity is penetrable for neutrons. Yet, in our microscopic simulations, we find successful vortex ring generation cases. For example, in the cases presented in Fig. 3, we denoted the vortex ring's appearance with circles. One can notice that the creation of a vortex ring occurred always when the velocity of a nucleus or impurity exceeds the Landau velocity $v_\mathrm{L}$. These vortices may be generated one by one, preventing the impurity to move faster than $v_\mathrm{crit}$. With the increase of the magnitude of the force $F$, the rate of vortex creation increases, which we may understand by noting that the rate of energy pumping into the system depends on the force $\sim \boldsymbol{F} \cdot \boldsymbol{v}$. The velocities for which we have detected the vortex nucleation are summarized in Fig. 7.

Exceeding Landau's velocity for the relative flow generated by impurity and the sea of neutrons is not the only criterion that must be fulfilled. Once the vortex is nucleated, it must propagate with a slower velocity than the impurity, to be able to detach. The velocity of a vortex ring in the hydrodynamic regime is given by [111]

$$v_{vr}(r) = \frac{1}{4\pi r}\frac{h}{2m_n}\left(\ln\frac{8r}{r_{core}} - \alpha\right), \quad (23)$$

where we used the fact that the circulation unit is given by $\Gamma = h/2m_n$. The coefficient $\alpha$ depends on the selected vortex core model; typically, $\alpha \approx 1/2$. It assumes that the vortex ring is a smooth circle. In reality, the line can be wiggled due to Kelvin wave excitations that modify the velocity of the vortex ring, which are also seen in our simulations. The hydrodynamic formula is derived for the limit of the vortex ring radius $r$, being much larger than the vortex core radius $r_{core}$. In Fig. 7 we show the velocity predicted by formula (23), where we use $\alpha = 1/2$, radius $r = R$ given by formula Eq. (16), and $r_{core} = \xi$. It turns out that this velocity is very close to the critical velocity $v_{crit}$, which provides another argument indicating that it is one of the important velocity scales for the construction of effective models.

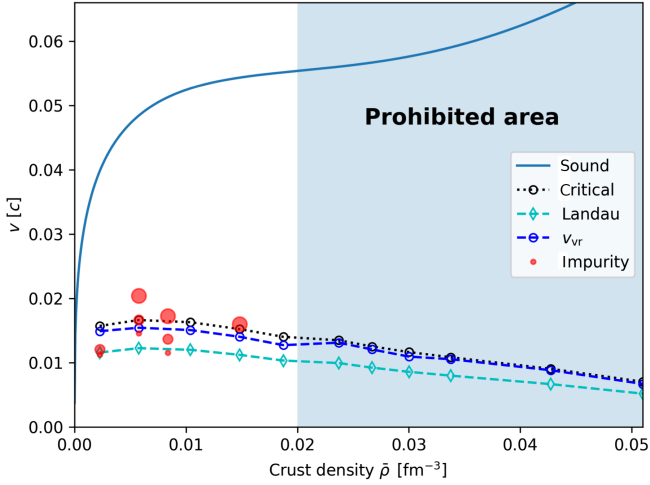

FIG. 7. The velocity scales in the system in units of the speed of light. The speed of sound (solid line) of the system is a few times larger than the Landau $v_L$ (dashed green) and critical $v_{crit}$ (dotted black) velocities. We plot the hydrodynamic velocity of a vortex ring $v_{vr}$ with the dashed blue line. By red dots, we denote the impurity velocity at which the vortex ring is created of three different sizes for different forces (MeV/fm): $F = 0.5$, 1, and 2, respectively. The vortex rings are not produced in the blue area on the right-hand side.

Systematic studies for various densities show that vortex rings are not always formed. We have not detected nucleation of vortices for densities above $\bar{\rho} > 0.02$ fm$^{-3}$. This threshold value coincides with densities at which the coherence $\xi$ becomes larger than the impurity size. The coherence length $\xi = \hbar k_{Fn}/(\pi\Delta_n m_n)$ of superfluid neutrons has a nontrivial behavior. The Fermi momentum $k_{Fn}$ is an increasing function of density $\rho_n$, while $\Delta_n$ has a maximum at $\rho_n \approx 0.017$ fm$^{-3}$. The minimum coherence length is at density $\rho_n \approx 0.006$ fm$^{-3}$. The size of the impurity depends on the density as well. In Fig. 7, we indicate by blue color the density range where the coherence length is slightly larger than the nucleus radius $\xi > 120\%R$. Clearly, in this regime, we do not detect the vortex nucleation process: One cannot generate a vortex of radius that is smaller than the size of the vortex core.

The numerical simulations demonstrate that vortex rings can be nucleated in the crust of a neutron star in the layers, where the average density $\bar{\rho}$ is below 0.02 fm$^{-3}$ provided the relative superfluid velocity exceeds (at least locally) Landau's velocity $v_L$. Studies of superfluid helium show that the injection of vortex rings is a very efficient way of generating quantum turbulence [112–115]. While our microscopic simulations cannot provide a definite answer on whether quantum turbulence can be present in neutron stars, they clearly show how quantum turbulence could develop at the smallest scale. This may have important implications for the global dynamics of neutron stars and the interpretation of sudden spin-ups seen as frequency glitches in some pulsars [116]. Although superfluidity is expected to play a major role, the actual triggering mechanism remains uncertain. Quantum turbulence could be one of them [117].

## VI. SUMMARY

To conclude, we have demonstrated that the quality of present DFT techniques combined with modern HPC solutions allows for the investigation of microscopic properties of nuclear systems and their quantum dynamics at the smallest scales relevant to neutron star crust. This is a critical step to construct global effective hydrodynamical models of neutron stars by averaging the local dynamics at smaller scales. Matter element exceeding the size of WS cells can be considered for both static and time-dependent HFB calculations without any symmetry restriction. In particular, time-dependent DFT may shed a new light on problems inaccessible so far through static approaches. Full 3D time evolutions without making any assumption on the weakness of external perturbations allow us to investigate a plethora of aspects of neutron star crusts. As an illustration of the possibilities offered by the toolkit, we have considered a simple numerical experiment in which a nuclear impurity is accelerated through a neutron superfluid medium by a constant force. Results of our simulations have provided new insight into the effective mass of the impurity, the characteristic velocity scales, the dissipative channels, and the mechanism for generation of topological defects. In particular, we have shown that low lying excitation modes, which are analogs of IGDR in atomic nuclei, can accompany the motion of impurity and therefore have to be included in any low energy description of neutron star crust.

Collecting such vast information has been possible by combining both static and time-dependent approaches: The system has been evolved from an initial configuration generated by self-consistent HFB calculations. The problem addressed here is only one from a large variety of questions, related to neutron stars, which can be investigated in a similar way. Among others, we can list the dynamics of quantum vortices in neutron star crust (proof of concepts were already demonstrated in Refs. [118,119]), pycnonuclear fusion processes, properties of exotic phases like nuclear pasta, and many others [120,121].

The number of open problems that are related to neutron stars is large. Collective effort within a unified approach is needed to foster progress. In particular, new functionals are being elaborated to fulfill the specific needs of time-dependent simulations of the kind presented here. Namely, the so-called family of functionals Brussels-Skyrme-on-a-grid (BSkG) is optimized toward applications involving calculations on 3D meshes [122–125], as presented in this paper. We strongly advocate for the further developments of a unified set of tools for the microscopic study of neutron stars in the future. This work aims at stimulating this effort and providing a software library in which other functionals can be easily implemented by making the code open source.

Reproducibility packs are provided in the dataset and supplement [97]. They provide the complete information needed to reproduce the results presented in this paper. Together with this paper, we make the associated toolkit publicly available.

## ACKNOWLEDGMENTS

D. P. acknowledges the hospitality from Université Libre de Bruxelles. We thank Nikolai Shchechilin for sharing cluster profiles data used for comparison. This work was financially supported by the (Polish) National Science Center Grants No. 2021/40/C/ST2/00072 (D. P.), No. 2021/43/B/ST2/01191 (P. M.), No. 2022/45/B/ST2/00358 (G. W.). This work was also supported by the Fonds de la Recherche Scientifique (Belgium) under Grant No. PDR T.004320 (N. C.). We acknowledge PRACE for awarding us access to resource Piz Daint based in Switzerland at Swiss National Supercomputing Centre (CSCS), decision No. 2021240031. We acknowledge Polish high-performance computing infrastructure PLGrid for awarding this project access to the LUMI supercomputer, owned by the EuroHPC Joint Undertaking, hosted by CSC (Finland) and the LUMI consortium through PLL/2022/03/016433.

D. P. and G. W. worked on construction of the `W-BSk Toolkit`; numerical calculations were performed by D. P. and A. Z.; D. P., A. Z., and G. W. analyzed the results. All authors contributed to research, planning, and interpretation of the results and manuscript writing.

## APPENDIX A: BRUSSELS-MONTREAL FUNCTIONALS

Here we provide the technical details concerning the type of functionals we are using, and the expression of the mean fields implemented into `W-BSk Toolkit`. Our code can handle semilocal functionals from the Brussels-Montreal family based on generalized Skyrme effective interactions with density-dependent $t_1$ and $t_2$ terms for the normal part [55], and a density-dependent contact interaction for the pairing part [57]. In this work, we have adopted the BSk31 functional, whose parametrization can be found in Ref. [30]. The expressions for the energy density and the mean fields in the absence of currents can be found in Ref. [55]. The additional terms depending on the currents were explicitly given in our previous work [84] for the special case of pure neutron matter. Here we provide expressions for arbitrary composition.

Below we introduce the isospin index $\iota = 0, 1$ for isoscalar and isovector quantities, respectively. Isoscalar quantities (also written without any subscript) are sums over neutrons and protons (e.g., $\rho_0 \equiv \rho_n + \rho_p$), while isovector quantities are differences between neutrons and protons (e.g., $\rho_1 \equiv \rho_n - \rho_p$). The different terms of the nuclear energy-density functional take the following forms [126]:

$$\mathcal{E}_\rho = \sum_\iota C_\iota^\rho[\rho]\rho_\iota^2, \tag{A1}$$

$$\mathcal{E}_\tau = \sum_\iota C_\iota^\tau[\rho](\rho_\iota \tau_\iota - \boldsymbol{j}_\iota^2), \tag{A2}$$

$$\mathcal{E}_{\Delta\rho} = \sum_\iota C_\iota^{\Delta\rho} \rho_\iota \Delta\rho_\iota + C_\iota^{\nabla\rho}[\rho]\Delta\rho_\iota. \tag{A3}$$

The coupling coefficients are related to the parameters of the generalized Skyrme interaction as

$$\begin{aligned}
C_0^\rho[\rho] &= \frac{3}{8}t_0 + \frac{3}{48}t_3\rho^\alpha, \\
C_1^\rho[\rho] &= -\frac{1}{4}t_0\left(\frac{1}{2}+x_0\right) - \frac{1}{24}t_3\left(\frac{1}{2}+x_3\right)\rho^\alpha, \\
C_0^\tau[\rho] &= \frac{3}{16}t_1 + \frac{1}{4}t_2\left(\frac{5}{4}+x_2\right) + \frac{3}{16}t_4\rho^\beta + \frac{1}{4}t_5\left(\frac{5}{4}+x_5\right)\rho^\gamma, \\
C_1^\tau[\rho] &= -\frac{1}{8}t_1\left(\frac{1}{2}+x_1\right) + \frac{1}{8}t_2\left(\frac{1}{2}+x_2\right) \\
&\quad -\frac{1}{8}t_4\rho^\beta\left(\frac{1}{2}+x_4\right) + \frac{1}{8}t_5\left(\frac{1}{2}+x_5\right)\rho^\gamma, \\
C_0^{\Delta\rho}[\rho] &= -\frac{9}{64}t_1 + \frac{1}{16}t_2\left(\frac{5}{4}+x_2\right) - \frac{3}{32}t_4\rho^\beta, \\
C_1^{\Delta\rho}[\rho] &= \frac{3}{32}t_1\left(\frac{1}{2}+x_1\right) + \frac{1}{32}t_2\left(\frac{1}{2}+x_2\right) \\
&\quad + \frac{1}{16}t_4\left(\frac{1}{2}+x_4\right)\rho^\beta, \tag{A4}
\end{aligned}$$

$$C_0^{\nabla\rho}[\rho] = \frac{3}{64}t_4\rho^\beta - \frac{1}{16}t_5\left(\frac{5}{4}+x_5\right)\rho^\gamma, \quad (A5)$$

$$C_1^{\nabla\rho}[\rho] = -\frac{1}{32}t_4\left(\frac{1}{2}+x_4\right)\rho^\beta - \frac{1}{32}t_5\left(\frac{1}{2}+x_5\right)\rho^\gamma. \quad (A6)$$

The pairing contribution to the energy-density functional is given by Eq. (9) of Ref. [30]:

$$\mathcal{E}_\pi = \frac{1}{4}f_n^\pm(v^{\pi n}(\rho_n,\rho_p) + \kappa_n|\nabla\rho_n|^2)|\nu_n|^2$$
$$+ \frac{1}{4}f_p^\pm(v^{\pi p}(\rho_n,\rho_p) + \kappa_p|\nabla\rho_p|^2)|\nu_p|^2, \quad (A7)$$

where $f_q^\pm \approx 1$. In our calculations, we set $f_q^\pm = 1$ and we drop the gradient terms. The pairing strengths $v^{\pi q}(\rho_n,\rho_p)$ were constructed so as to reproduce exactly the $^1S_0$ pairing gaps obtained in Ref. [38] in pure neutron matter and symmetric nuclear matter from extended Brueckner-Hartree-Fock calculations including medium polarization and self-energy effects. We adopt the analytical approximation given in Ref. [58]:

$$v^{\pi q} = -\frac{8\pi^2}{\sqrt{\mu_q}}B_q^{3/2}\left[2\ln\left(\frac{2\mu_q}{|\Delta_q|}\right) + \Lambda\left(\frac{E_{\text{cut}}}{\mu_q}\right)\right]^{-1}, \quad (A8)$$

where $\mu_q$ is the chemical potential, $E_{\text{cut}}$ is the cutoff energy, and

$$\Lambda(x) = \ln(16x) + 2\sqrt{1+x} - 2\ln(1+\sqrt{1+x}) - 4. \quad (A9)$$

The single-particle Hamiltonian $h_q$ can be obtained by varying the energy-density functional $\mathcal{E}$ with respect to the various densities and currents Eqs. (4)–(7), and reads

$$h_q = U_q^\rho + U_q^{\Delta\rho} + U_q^\tau + U_q^\pi - \nabla B_q \nabla - \frac{i}{2}\{A_q,\nabla\}. \quad (A10)$$

The first four terms are scalar potentials coming from variation over the density $\rho_q$:

$$U_q^\rho = \frac{\partial\mathcal{E}_\rho}{\partial\rho_q}, \quad (A11)$$

$$U_q^\tau = \frac{\partial\mathcal{E}_\tau}{\partial\rho_q}, \quad (A12)$$

$$U_q^{\Delta\rho} = \frac{\partial\mathcal{E}_{\Delta\rho}}{\partial\rho_q} - \nabla\cdot\frac{\partial\mathcal{E}_{\Delta\rho}}{\partial(\nabla\rho_q)} + \Delta\frac{\partial\mathcal{E}_{\Delta\rho}}{\partial(\Delta\rho_q)}, \quad (A13)$$

$$U_q^\pi = \frac{\partial\mathcal{E}_\pi}{\partial\rho_q} - \nabla\cdot\frac{\partial\mathcal{E}_\pi}{\partial(\nabla\rho_q)}. \quad (A14)$$

We checked that $U_q^\pi$ is very small compared to other terms; therefore, we neglect it in our calculations. The field $B_q$ arises from the dependence of the functional on the kinetic density $\tau_q$:

$$B_q = \frac{\hbar^2}{2m_q} + \frac{\partial\mathcal{E}_\tau}{\partial\tau_q}. \quad (A15)$$

The last term is a vector potential induced by the presence of a current $j_q$:

$$A_q = \frac{\partial\mathcal{E}_\tau}{\partial j_q}. \quad (A16)$$

The various mean fields are expressible as

$$B_q = \frac{\hbar^2}{2m_q} + C_0^\tau\rho + C_1^\tau(\rho_q - \rho_{q'}), \quad (A17)$$

$$A_q = -2C_0^\tau j - 2C_1^\tau(j_q - j_{q'}), \quad (A18)$$

$$U_q^\rho = \frac{dC_0^\rho}{d\rho}\rho_n^2 + \frac{dC_1^\rho}{d\rho}(\rho_q - \rho_{q'})^2 + 2\rho C_0^\rho + 2(\rho_q - \rho_{q'})C_1^\rho, \quad (A19)$$

$$U_q^\tau = C_0^\tau\tau + C_1^\tau(\tau_q - \tau_{q'}) + \frac{dC_0^\tau}{d\rho}(\rho\tau - j^2)$$
$$+ \frac{dC_1^\tau}{d\rho}[(\rho_q - \rho_{q'})(\tau_q - \tau_{q'}) - (j_q - j_{q'})^2], \quad (A20)$$

$$U_q^{\Delta\rho} = 2C_0^{\Delta\rho}\Delta\rho + 2C_1^{\Delta\rho}(\Delta\rho_q - \Delta\rho_{q'}) + \frac{dC_0^{\nabla\rho}}{d\rho}\Delta\rho$$
$$+ \frac{dC_1^{\nabla\rho}}{d\rho}(\Delta\rho_q - \Delta\rho_{q'}) + \nabla\cdot\left(\frac{dC_0^{\nabla\rho}}{d\rho}\nabla\rho\right)$$
$$+ \nabla\cdot\left(\frac{dC_0^{\nabla\rho}}{d\rho}\nabla(\rho_q - \rho_{q'})\right). \quad (A21)$$

Here $q'$ means the complementary nucleon species to $q$ (if $q = n$, $q' = p$ and vice versa).

## APPENDIX B: PROTOCOL FOR TURNING ON THE ELECTRIC FIELD

The static solutions are generated with no external potentials and are used as starting points for time-dependent considerations. To initialize dynamics, we turn on the electric field in the following way:

$$E_z(t) = \begin{cases} 0 & \text{if } t < t_1 \\ E_z s(t - t_1, t_2 - t_1) & \text{if } t_1 < t < t_2 \\ E_z & \text{if } t_2 < t, \end{cases} \quad (B1)$$

where $t$ is time. $t_1$, $t_2$ denote moments of time when the electric field starts to rise smoothly and saturates at the final value, respectively. The switching function $s$ that models the rising of the field is

$$s(t,\Delta t) = \frac{1}{2}\left\{1 + \tanh\left[\tan\left(\pi\frac{t}{\Delta t} - \frac{\pi}{2}\right)\right]\right\}. \quad (B2)$$

We used $\Delta t = t_2 - t_1 = 10$ fm/$c$ in our calculations.

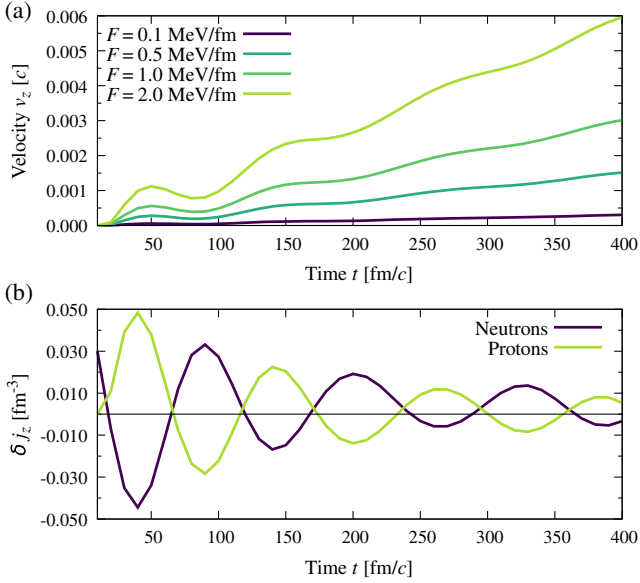

FIG. 8. (a) Enlargement for short time from Fig. 3(d). One can observe damped oscillations on top of a linear increase in velocity. (b) Neutron and proton currents along z direction for $F = 2$ MeV/fm. We removed a linear increase of the current seen in (a) that is not relevant for the effect. See the main text for details.

## APPENDIX C: GIANT DIPOLE RESONANCE IN NEUTRON MATTER

If one looks closer at the first moments of our simulation, the increase of velocity $v_z(t)$ is not linear. For example, in Fig. 8(a), we present an enlargement of the initial stage of motion from Fig. 3(d). On top of linear behavior, there are damped oscillations. These oscillations are generated as a side effect of our numerical setup; however, they have a well-defined physical origin.

The oscillations are caused by the fact that we "turn on" the electric field $E_z$, Eq. (B1), once the static solution, representing impurity at rest, is generated, although we do not do it instantaneously but smoothly increase the interaction during a finite but short time $\Delta t$. The smaller the period $\Delta t$, or the larger the force $F$, the larger becomes the amplitude of oscillations. Since only protons couple to the electric potential, they start to move as first. Next, they begin to drag neutrons bound to them through nuclear forces. Such a scenario gives rise to an excitation of an analog of IGDR mode in our system. It is the basic excitation mode in nuclear systems, where protons vibrate (practically harmonically) against neutrons [127,128]. IGDR is not dissipationless mode, resulting in its finite width due to coupling to more complex nuclear configurations or to the continuum [129]. As evidence that indeed we induce IGDR, we plot the currents for neutrons and protons in Fig. 8(b), but with subtracted constant linear flow of the whole nucleus. We see that protons and neutrons are out of phase. Moreover, we can see from the figure that the period of oscillations is approximately $T \approx 100$ fm/$c$. This is consistent with the phenomenological formula for the IGDR frequency $\Omega_{\mathrm{IGDR}} = 0.39 A^{-1/3} c/\mathrm{fm}$ [130], even though it has been designed for finite-size nuclei in vacuum. The formula with $A = 140$ provides the oscillation period to be about 84 fm/$c$ which is of the same order as observed.

## APPENDIX D: PROPERTIES OF THE INNER CRUST

We provide Table I with a summary of characteristic quantities for considered cases.

TABLE I. Quantities extracted from simulations for each density of the inner crust $\bar{\rho}$: $\rho_{\mathrm{B}n}$, bulk density of neutrons; $\Delta_n$, pairing energy of neutrons; $k_{\mathrm{F}n}$, wave vector calculated for bulk density of neutrons; $\varepsilon_{\mathrm{F}n}$, neutron Fermi energy; $\varepsilon_{\mathrm{F}n}^*$, neutron Fermi energy calculated with bare mass; $N_n$, number of neutrons; $\xi$, coherence length; $R$, radius of impurity; $M_{\mathrm{eff}}$, effective mass of impurity.

| $\bar{\rho}$ (fm$^{-3}$) | $\rho_{\mathrm{B}n}$ (fm$^{-3}$) | $\Delta_n$ (MeV) | $k_{\mathrm{F}n}$ (fm$^{-1}$) | $\varepsilon_{\mathrm{F}n}$ (MeV) | $\varepsilon_{\mathrm{F}n}^*$ (MeV) | $N_n$ | $\xi$ (fm) | $R$ (fm) | $M_{\mathrm{eff}}$ ($m_n$) |
|---|---|---|---|---|---|---|---|---|---|
| 0.0023 | 0.0016 | 0.826 | 0.363 | 2.723 | 2.694 | 376.6 | 5.79 | 5.32 | 150.75 ± 1.8 |
| 0.0058 | 0.0045 | 1.236 | 0.512 | 5.436 | 5.301 | 934.4 | 5.47 | 5.21 | 139.30 ± 1.0 |
| 0.0104 | 0.0084 | 1.483 | 0.628 | 8.165 | 7.840 | 1696.8 | 5.58 | 5.64 | 164.70 ± 3.1 |
| 0.0148 | 0.0120 | 1.562 | 0.707 | 10.37 | 9.839 | 2385.3 | 5.98 | 5.79 | 155.25 ± 2.9 |
| 0.0187 | 0.0152 | 1.557 | 0.766 | 12.15 | 11.421 | 3032.5 | 6.49 | 6.21 | 174.45 ± 4.1 |
| 0.0237 | 0.0193 | 1.621 | 0.829 | 14.24 | 13.248 | 3789.3 | 6.75 | 5.28 | 168.55 ± 6.3 |
| 0.0267 | 0.0217 | 1.566 | 0.863 | 15.42 | 14.268 | 4258.0 | 7.27 | 5.55 | 168.80 ± 7.1 |
| 0.0300 | 0.0244 | 1.514 | 0.898 | 16.70 | 15.368 | 4847.1 | 7.82 | 7.46 | 171.85 ± 8.3 |
| 0.0338 | 0.0276 | 1.467 | 0.935 | 18.10 | 16.558 | 5430.9 | 8.40 | 7.27 | 166.15 ± 8.3 |
| 0.0428 | 0.0351 | 1.327 | 1.013 | 21.28 | 19.260 | 6925.0 | 10.1 | 8.71 | 150.50 ± 8.8 |
| 0.0510 | 0.0422 | 1.097 | 1.077 | 24.05 | 21.618 | 8070.6 | 13.0 | 9.03 | 150.60 ± 8.8 |